# Surface Temperature Trend Estimation over 12 Sites in Guinea Using 57 Years of Ground-Based Data

**René Tato Loua [1,2,3,\*], Hassan Bencherif [1,4], Nelson Bègue [1], Nkanyiso Mbatha [5], Thierry Portafaix [1], Alain Hauchecorne [6], Venkataraman Sivakumar [4] and Zoumana Bamba [2]**

1. Laboratoire de l'Atmosphère et des Cyclones, UMR 8105, CNRS, Université de La Réunion, Météo-France, 97490 Réunion, France; hassan.bencherif@univ-reunion.fr (H.B.); nelson.begue@univ-reunion.fr (N.B.); thierry.portafaix@univ-reunion.fr (T.P.)
2. Centre de Recherche Scientifique de Conakry Rogbane, Conakry 1615, Guinea; vanzoumana@yahoo.fr
3. Direction Nationale de la Météorologie de Guinée, Conakry 566, Guinea
4. School of Chemistry and Physics, University of KwaZulu Natal, Durban 4000, South Africa; Venkataramans@ukzn.ac.za
5. Department of Geography, University of Zululand, KwaDlangezwa 3886, South Africa; mbathanb@unizulu.ac.za
6. Laboratoire Atmosphère, Milieux, Observations Spatiales/Institut Pierre-Simon-Laplace, UVSQ Université Paris-Saclay, Sorbonne Université, CNRS, 78280 Guyancourt, France; Alain.Hauchecorne@latmos.ipsl.fr
\*: Correspondence: rene-tato.loua@univ-reunion.fr

**Abstract:** Trend-Run model was performed to estimate the trend in surface temperatures recorded at 12 sites in Guinea from 1960 to 2016 and to examine the contribution of each climate forcing. The coefficient of determination ($R^2$) calculated varies between 0.60 and 0.90, it provides total information about the simulation capability of the model. The decadal trend values also calculated show an upward trend (between 0.04 °C ± 0.06 °C decade$^{-1}$ and 0.21 °C ± 0.06 °C decade$^{-1}$). In addition, forcings' contributions were quantified, and the annual oscillation (AO) contribution is higher for most of the stations, followed by semiannual oscillation (SAO). Among the forcings, the tropical Northern Atlantic (TNA) contribution is greater than that of the sunspot number (SSN), Niño3.4 and Atlantic Niño (AN). Moreover, the Mann-Kendall test revealed a positive significant trend for all stations except at the Macenta site. Additionally, with sequential Mann-Kendall test, trend turning points were found only for the stations of Mamou, Koundara and Macenta at different dates. The temperature anomalies depict warming episodes (1970s, 1980s, 1984 and 1990s). Since then, the temperature is consistently increasing over the country. A significant warming has been shown, which might be further investigated using these models with additional contributing factors.

## 1. Introduction

In the 20th century (1923–1950), warming occurred mainly in the Northern Hemisphere mid- and high-latitudes, whereas, in the most recent decades (1977–2014), warming was more spatially extensive across the global land surface [1]. According to the Intergovernmental Panel on Climate Change (IPCC) Special Report [2], anthropogenic warming was reported to have already exceeded preindustrial levels by about 1°C. That is why, during the last decade, and particularly since the mid-2000s, the scientific community has begun to engage seriously with the issue of climate change and its implications for developing countries [3]. Since then, the study of the variability of climatic parameters is one of the major concerns of the scientific community [3–5]. That is why it is noted that the findings of researches are useful tools to assist decision-makers in the context of climate change [6].

Some developing countries are limited in their ability to cope with the problems caused by climate change mainly because of technical, technological and economic shortcomings. Recently, Barry et al. [7] highlighted that the lack of long-term and high-quality data availability was seen as a limiting factor in conducting a study in the framework of observational analyses on changing climate extremes over the West Africa region. Therefore, the African continent is likely to face extreme and widespread droughts now and in the future. This evident challenge is likely to aggravate due to slow progress in drought risk management, increased population and demand for water and degradation of land and environment [8]. Moreover, there has been a pattern of continuing aridity in West Africa since the late 1960s [9]. Given these facts, and with the resurgence of climate change impacts, the international scientific community has a special focus on climate research.

It is important to retain the ground-based surface temperature in the analysis of climate variability. For instance, from a study on a temperature trend analysis in South Africa, it was highlighted that temperature can, directly and indirectly, impact the livelihood of inhabitants of the country and the natural environment as a whole [10]. Temperature has been the subject of several studies on different atmospheric layers according to various research interests [3,11–22]. Ahmed et al. [23] used statistical analyses (Fourier test analysis, $t$-test analysis and Mann-Kendall test analysis) for determining the trend in the annual and seasonal temperature time-series at fifteen weather stations within the Ontario Great Lakes Basin for the period 1941–2005.

The term "trend" is widely used in the vernacular to mean general direction or tendency [24]. In fact, the trend is often the first thing to detect in the analysis of a time-series. It is the general orientation of a series of observations up or down over a long period. Since trend detection in temperature and precipitation time-series is one of the interesting research areas in climatology [25], the trend and periodicity analysis of temperature time-series is of great importance because of the effect of global climate change, and most time-series patterns can be described by trend and seasonality [23]. Thus, the analytical method of trend estimation is therefore based not only on theoretical concepts but, also, on choices: first, the determination of the trend model (a linear or exponential function of time, for example) and an additive or multiplicative decomposition scheme that allows the seasonal component to be isolated. Thus, there are several methods of trend analysis, and the choice of the desired method depends on the study context. For instance, the Mann-Kendall (MK) trend test [26,27], multilinear regression [28,29] and Trend-Run model [14] were used in order to investigate different atmospheric parameters [19,30,31].

Moreover, it should be noted that the variation of the surface temperature remains dependent on certain non-meteorological factors such as warm bias in nighttime minimum temperatures, poor siting of the instrumentation and effect of wind, as well as surface atmospheric water vapor content on temperature trends, the quantification of uncertainties in the homogenization of surface temperature data and the influence of land use/land cover changes [32]. Especially, land cover changes (LCCs) is a factor that may rapidly change local climates. Some research works have analyzed these LCCs and their impacts on the climate, such as that of Mahmood et al. [33], who studied the impacts of notable types of land cover changes, including agriculture, deforestation and afforestation, desertification and urbanization on the climate. In addition, the agricultural expansion influence on the climate over the full troposphere in Northeast China from 1982 to 2010 was studied by He et al. [34].

Based on our knowledge of understanding, a very few studies on temperature variability have been conducted in some sites in Guinea [18,19], but a trend analysis using this specific method has not yet been done. In this context, we applied the Trend-Run model and Mann-Kendall trend test for a trend analysis of temperatures over Guinean regions. Moreover, it is the first time these methods are going to be evaluated in this region with its specific environmental and geographical contexts. The main objective of this study was based (1) on a climatological variability analysis of temperature over the country, (2) on the trend estimates of temperatures recorded at 12 weather stations in Guinea and (3) on quantifying the models' performances and climate forcing. In general, this manuscript is of interest as a decision support tool, as it will not only improve our understanding of the spatio-temporal variability of surface temperatures in Guinea but will also provide new knowledge on the evolution of the observed trends.

In the following sections, brief details of the study area are given in Section 2. The details of the dataset used in the study, as well as the different methods adopted for trend analysis, are presented in Section 3. Before concluding the study in Section 5, the results obtained from different methods deployed for trend analysis are shown and discussed in Section 4.

## 2. Materials and Methods

### 2.1. Study Site Description

The Republic of Guinea is located at the Southwestern region of West Africa covering a surface of 245,857 km². It is a coastal country with 300 km of Atlantic coastline, midway between the equator and the tropic of cancer, between 7°05 and 12°51 N and 7°30 and 15°10 W. It is limited in the East by the Ivory Coast, in the North-eastern part by Mali, in the North by Senegal, in the North-western part by Guinea Bissau, in the South by Libera and Sierra Leone and in the West by the Atlantic Ocean (see Figure 1). Additionally, this region is subjected to the alternating action of the North-Easterly wind (the harmattan) and the South-Westerly air mass (the monsoon), which are associated with the dry (winter) and wet seasons (summer), respectively.

Due to the presence of distinct geographical and climatic conditions, Guinea is subdivided into 4 natural regions: (1) Lower Guinea, which occupies the entire coastal strip of the country with mangrove forests and lowlands, (2) Middle Guinea covers the mountainous and wooded savannah area where the Fouta Djallon massif is located, (3) Upper Guinea is an upland area and savannah covered by the largest watershed area of the country (Niger's watershed) and (4) Forest Guinea, which occupies the Guinean Ridge massif; it is the region of both dense forests and mountains, with also a dense river network. It should be noted that Guinea is one of the most watered countries in the West African region; the annual average rainfall varies from 1300 mm (Northern part) to 4000 mm (coastal region) and peaks throughout July-August. It is characterized by two seasons, a dry season with about 3 months in duration (in the North) and a rainy season with about 8–9 months in duration (in the Southeast). Moreover, as mentioned above, it has a very dense river network, with about 1161 rivers that spring from the two mountainous massifs (the Fouta Djallon and the Guinean Ridge).

Therefore, in this study, we take into account this geographical distribution in the use of the datasets from 12 weather stations distributed in the 4 natural regions (3 stations by region). Images of some stations are shown in Appendix A.

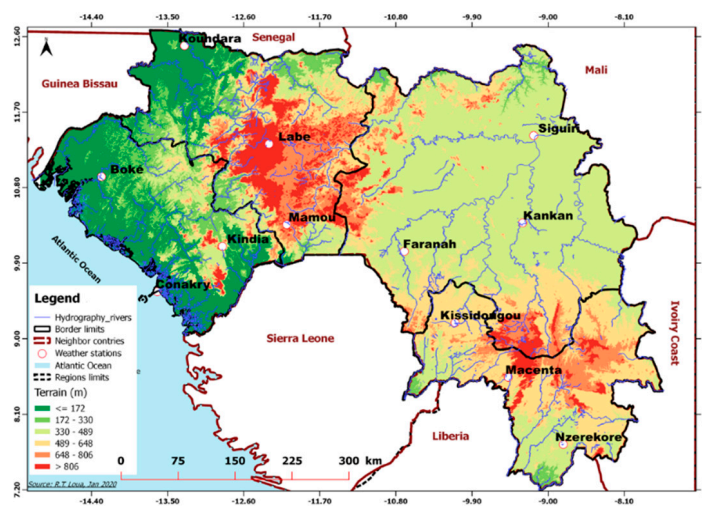

**Figure 1.** Geographical location of 12 weather stations used in the present study (circle symbols) along with elevation (in meters in ground color according to the legend) in the four natural regions of Guinea. Edited by Loua R.T. 02/2020.

*2.2. Data*

2.2.1. Ground-based Data

Identification of climate trends often requires large regional or national scales and long time periods, as climate variability can mask trends in local areas and over short timeframes. In the present study, the Trend-Run model was adapted for the National Meteorological Service of Guinea (NMSG) surface temperature datasets. The monthly temperature data were accessed from the NMSG database for the 12 stations covering the whole country (see Table 1). NMSG is a public service working in the administrative domain of the ministry of transportations. They are responsible for the collection, quality-checking of meteorological observations and weather forecasting. The observation network of this service is faced not only with a low density of stations number but, also, with the obsolescence and lack of modern efficient equipment. Therefore, 12 weather stations with reliable data among many others are considered in this study. Remotely sensed data is used as complementary data for the spatial analysis of temperature.

**Table 1.** The corresponding values of geographical information for the 12 weather stations used in the study along with availability of the temperature records in: Lower-Guinea (Conakry, Boke and Kindia); Middle-Guinea (Labe, Mamou and Koundara); Upper-Guinea (Kankan, Siguiri and Faranah) and Forest Guinea (N'zerekore, Macenta and Kissidougou) starting and ending years with available data for the surface monthly averaged temperatures in Guinea.

| Station | Long (°) | Lat (°) | Alt (m) | Starting Year | Ending Year | Available Data (monthly/year) |
|---|---|---|---|---|---|---|
| *Lower-Guinea: lowland area, very watered and high temperatures because of mountains, urban and coastal effects.* | | | | | | |
| Conakry | -13.37 | 9.34 | 46 | 1960 | 2016 | 684/57 |
| Boke | -14.18 | 10.56 | 69 | 1960 | 2016 | 684/57 |
| Kindia | -12.86 | 10.04 | 458 | 1960 | 2016 | 684/57 |
| *Middle-Guinea: mountainous area, both lower temperature (Labe and Mamou) and higher temperature (Koundara).* | | | | | | |
| Labe | -12.29 | 11.19 | 1050 | 1960 | 2016 | 684/57 |
| Mamou | -10.08 | 10.38 | 782 | 1960 | 2016 | 684/57 |
| Koundara | -13.31 | 12.34 | 90 | 1975 | 2016 | 504/42 |
| *Upper-Guinea: Upland area, high temperatures and lower rainfall.* | | | | | | |
| Kankan | -9.55 | 10.12 | 376 | 1960 | 2016 | 684/57 |
| Siguiri | -9.37 | 11.74 | 361 | 1960 | 2000 | 492/41 |
| Faranah | -10.8 | 10.26 | 358 | 1968 | 1997 | 360/30 |
| *Forest Guinea: Forest and mountainous region; very watered with longest length; more rivers.* | | | | | | |
| N'zerekore | -8.83 | 7.75 | 467 | 1960 | 2016 | 684/57 |
| Macenta | -9.28 | 8.32 | 542 | 1960 | 2000 | 492/41 |
| Kissidougou | -10.11 | 9.19 | 524 | 1974 | 2009 | 432/36 |

2.2.2. MERRA 2 Model Data

In the present study, the Modern Era Retrospective-Analysis for Research and Applications (MERRA 2) monthly surface temperature data is used. MERRA2 data is an area averaged at 2 meter of air temperature in each box at the spatial resolution of 0.50x0.625 with both dimensions along the latitude and longitude respectively provided in degrees, covering the period from 1980 to present. MERRA 2 time-series is available on the website: https://giovanni.gsfc.nasa.gov/giovanni/. The period from 1980 to 2016 is considered for study. Pearson method was employed to calculate the correlations [35] between the MERRA 2 data and the ground-based data. There is a strong positive correlation between the remotely sensed data and ground-based data at each site, with the r values that lie between 0.719 and 0.955 (see Figure 2). The good correlation found between the two datasets verifies the validation of the MERRA 2 data we used for the spatial analysis; the results are shown in the Section 3.3.

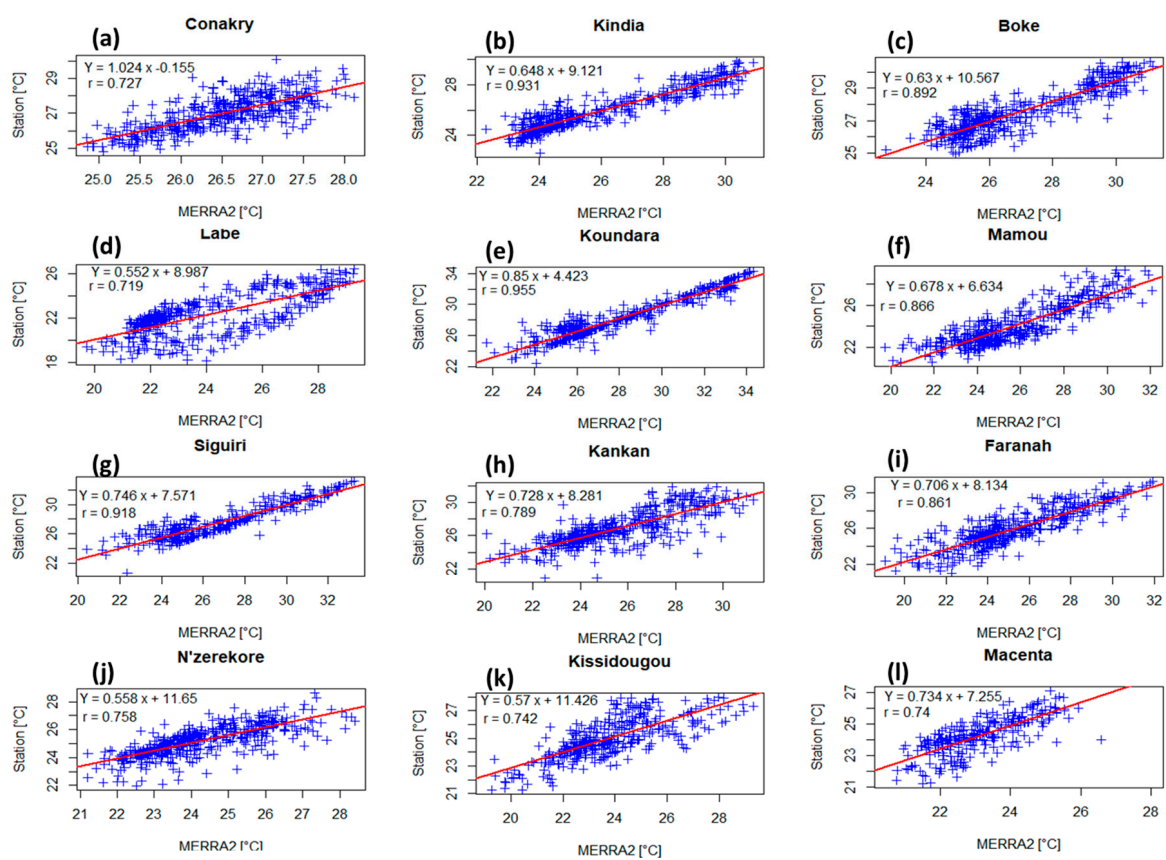

**Figure 2.** Pearson correlation between ground-based data and Modern Era Retrospective-Analysis for Research and Applications (MERRA 2) temperature data for the 12 sites (**a–l**) for the period from 1980 to 2016.

2.2.3. Climate Forcings

Atlantic Intertropical Convergence Zone (ITCZ) is controlled by changes in the inter-hemispheric sea surface temperature (SST) gradient [36]. Since the variability of the climate system in the tropical Atlantic shows an inherently atmosphere-ocean coupled behaviour [37], significant fluctuations in tropical SST are important for the climate of the tropical Atlantic and the surrounding land regions [38,39]. Then, the Trend-Run model further used explanatory variables. The application of this model on our study area might be based on the climate forcings' Tropical Norther Atlantic (TNA), Atlantic Meridional Mode (AMM), Atlantic Niño (AN), El Niño 3.4 (Niño 3.4) and Sun Spot Number (SSN). The temporal evolutions of these parameters are shown in Figure 3.

The variation of Atlantic SST influences some climate parameters in the surrounding regions. For instance, Enfield [40] highlighted that the rainfall variability in Eastern Caribbean and North-Eastern Brazil is associated with the TNA index variation. From the same source, the tropical SST variability was represented with two simple area indices north and south of the ITCZ corresponding to the tropical Northern Atlantic (TNA) and tropical Southern Atlantic (TSA), respectively. The TNA index is defined by Enfield et al [41] to be the average of the monthly data (SSTA) over the rectangular region 5°N-25°N, 55°W-15°W.

The AMM consists of a north-south SST gradient that exhibits opposite SST anomalies on either side of the mean position of the ITCZ [42] and operates on both decadal and inter-annual time scales [43]. This monthly SSTA index positive phase is related to a reduction of wind shear, increased SST and decreased pressure over the main development region defined as 20°-60°E, 10°-20°N [44]. Our study area is concerned by the AMM-related SST index for the area 32°N-21°S and 74°W to the West African coastline, which results from the calculation based on a maximum covariance analysis (MCA) [45], and its generation mechanism is designed by Amaya et al. [46] as the wind-evaporation-SST (WES) feedback. The positive/negative phase corresponds to positive/negative SST anomaly in the tropical North Atlantic and the opposite in the tropical South Atlantic [37].

The Atlantic equatorial mode is a quasiperiodic inter-annual climate pattern of the equatorial Atlantic Ocean characterized by a dominant mode of year-to-year variability that results in alternating warming and cooling episodes of the sea surface temperature accompanied by changes in atmospheric circulation [47]. Due to its close similarity with the El Niño-Southern oscillation, it is often called the Atlantic Niño-ATL3 [48,49] or Atlantic zonal mode [37]. Atlantic Niño is known to be strongly related to atmospheric climate anomalies, and it has an important implication for climate prediction, especially in African countries bordering the Gulf of Guinea. It has been defined by Zebiak [22] to be the area average of the detrended SST anomaly over the region 3° N-3° S, 0-20° W. While the low SST that occurs during June-July-August (JJA) in the Gulf of Guinea gives place to the Atlantic cold tongue (ACT) [50].

Barry et al. [7] highlighted the usefulness of understanding the influence of the El Niño southern oscillation (ENSO) on climate extremes variability over the West African region. The number of the Niño 1, 2, 3 and 4 regions correspond with the label assigned to ship tracks that cross these regions. The Niño 3.4 index (5° N-5° S, 150° W-90° W) is one of the more commonly used indices to define El Niño and La Niña events. The Niño 3.4 anomalies may be thought of as representing the average equatorial SST across the Pacific, from about the dateline to the South American coast. We use this index in the present study because of its known teleconnection with the Atlantic Ocean in order to identify its possible contribution in temperature variability over Guinea. The Niño 3.4 index typically uses a 5-month running mean, and El Niño or La Niña events are defined when the Niño 3.4 SST exceeds ± 0.4°C for a period of 6 months or more [51].

Since the solar energy flux is the principal source of energy for the Earth's surface, a solar flux variation seems to be considered as a possible source of climate variation. Many studies have been driven on by the link between solar irradiance and the Earth's surface temperature—for example, Zerbo et al. [52–55].

The Atlantic meridional mode (AMM) data is available from the University of Wisconsin website through the link: http://www.oas.wisc.edu/~dvimont/MModes/Data.html. The 10.7-cm solar flux (given in solar flux units: sfu = $10^{-22}$W m$^{-2}$ Hz$^{-1}$), the tropical North Atlantic (TNA), the Atlantic Niño (AN) and the Niño3.4 indices data are provided by the National Oceanic and Atmospheric Administration (https://www.esrl.noaa.gov/psd/).

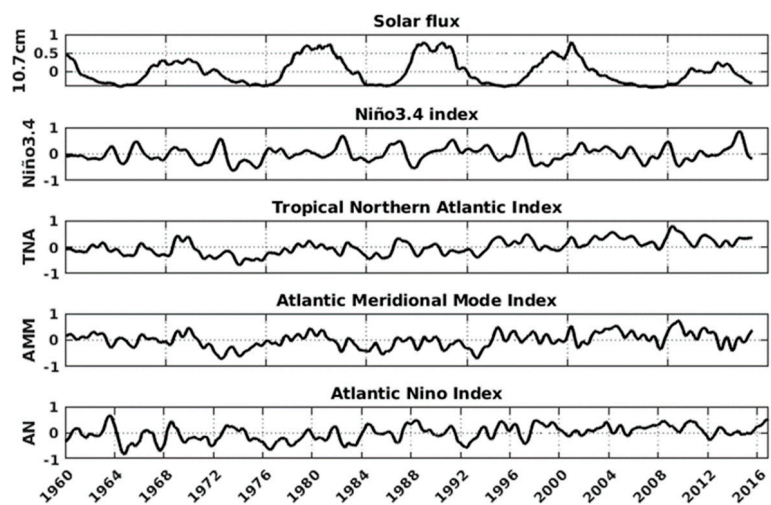

**Figure 3.** Temporal evolution of normalized climate indices: Sun Spot Number, El Niño 3.4, Tropical Northern Atlantic, Atlantic Meridional Mode and Atlantic Niño, from 1960 to 2016.

*2.3. Methods*

2.3.1. Trend-Run Model

Trend-Run is a statistical model from which the analyses are based on a linear regression fitting. It is adapted from the AMOUNTS (Adaptive MOdel UNambiguous Trend Survey) and AMOUNTSO3 models, developed for ozone and temperature trend assessments [55–58]. Trend-Run model was adapted and used at the University of Reunion Island for temperature and trend estimation in the southern subtropical upper troposphere-lower stratosphere (UTLS) [14]. Subsequently, this model has been used as a trend analysis tool in several studies in the Indian Ocean Basin [10,12,59,60].

This model was based on the principle of breaking down the variation of a time-series Y(t) into the sum of different parameters that explain the variation of Y(t):

$$Y(z,t) = c_1 SAO(z,t) + c_2 AO(z,t) + c_3(z) QBO(z_{40},t) + c_4(z) ENSO(t) + c_5(z) SSN(t) + \varepsilon(z,t), \quad (1)$$

where $\varepsilon$ is the residual term, assumed to consist of trend and noise.

When the coefficients $c_i$ (*i*=1 to 5) are calculated, the corresponding parameters are removed from the studied geophysical signal Y(t). The model then applies the least-squares method in order to minimize the sum of the residual squares and to determine the parameters coefficients **$c_i$**. Regarding the trend, it is parameterized as linear:

$$Trend(t) = \alpha 0 + \alpha 1.t, \quad (2)$$

where t denotes the time range, $\alpha 0$ is a constant and $\alpha 1$ is the slope of the Trend(t) line that estimates the trend over the time scale.

In its initial version, as used by Bencherif et al. [14], the Trend-Run model considers the main forcing, i.e., annual and semi-annual cycles, QBO (quasi-biennial oscillation), ENSO (El-Niño southern oscillation) and the 11-year solar cycle (sunspot number).

As we know that the Trend-Run model is an additive model based on a linear regression fitting, an important factor is that the dependent parameter (temperature) and nondependent parameters (forcings) are all quantitative variables. Thus, AO and SAO have been taken into account and, for their phase construction, a sinusoidal function is used, as detailed in the work of Toihir et al. [60] and Sivakumar et al. [61].

In our case, and in order to examine the existing link between internal variation, regional climate forcings and surface temperature variation, as well as the surface temperature trend estimation over Guinea, the Trend-Run model has been modified by introducing different other parameters. It is noteworthy that this model works well if the explanatory variables are orthogonal to each other, because multicollinearity occurs when independent variables in a regression model are correlated. The correlation increases the standard errors of the coefficients. Increased standard errors in turn means that coefficients of some independent variables may be found not to be significantly different from 0. That makes some variables statistically insignificant when they should be significant. It can cause problems when we fit the model and interpret the results. For the present study, it was previously mentioned that there is a strong correlation between the TNA and AMM [19]. Indeed, a simple correlation analysis between the five forcings aforementioned is depicted in Figure 4. Statistically, the TNA and AMM are strongly correlated (coefficient of correlation = 0.8); then, the AMM is excluded in this analysis.

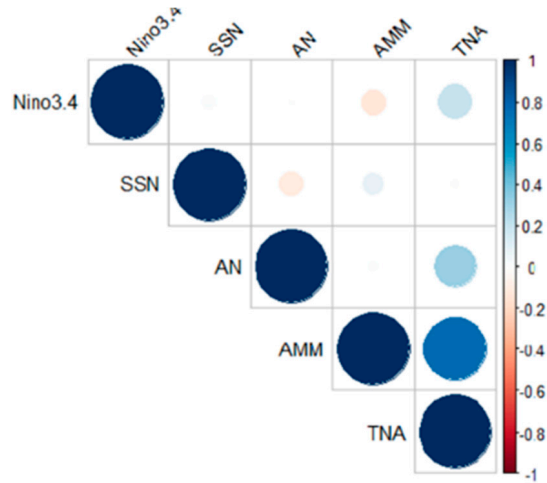

**Figure 4.** Positive correlations are displayed in blue and negative correlations in red. The intensity of the color and the size of the circles are proportional to the correlation coefficients. To the right of the correlogram, the color legend shows the correlation coefficients and the corresponding colours.

Inter-annual climate variability in the tropical Atlantic manifests itself in two distinct modes. The first is the equatorial or Atlantic Niño mode, and the second is the meridional or inter-hemispheric mode. TNA and AMM are meridional modes, since the TNA forcing seems to be more important than the AMM forcing we used as the explanatory variable to perform the Trend-Run model: AO, SAO, TNA, Niño3.4, AN and SSN, which are supposed to contribute to the variations of temperature in this area. The time-series indices of each of these parameters were firstly normalized, and then, the model equation in our study therefore becomes:

$$Y(z,t) = c_1 AO(z,t) + c_2 SAO(z,t) + c_3(z) TNA(t) + c_4(z) AN(t) + c_5(z) Niño3.4(z,t) + c_6(z) SSN(t) + \mathcal{E}(z,t), \quad (3)$$

where $c_i$ ($i$ = **1 to 6**) represented the various considered atmospheric force coefficients.

The coefficient of determination ($R^2$) is a vital coefficient that indicates the proportion of the total variation in temperature in time, explained by the Trend-Run. Its value is close to unity when the model explains almost all the variability of the signal and close to zero in the opposite case, and the sum of all forcings' contributions need to be close to $R^2$. The decadal trend values in degrees Celsius (°C) were also calculated to see if there was any change in the temperature increase or decrease over time [10].

2.3.2. Mann-Kendall Tests

The Mann-Kendall monotonic and sequential trend tests were performed. The MK trend test [62,63] is a nonparametric test which is used to statistically analyse data for consistently increasing (decreasing) trends (monotonic) in the variable through time. It was highlighted by Hirsch, Slack and Smith [64] that the MK test is best viewed as an exploratory analysis and is most appropriately used to identify stations where changes are significant or of large magnitude and to quantify these findings. The World Meteorological Organization (WMO) has also suggested using the Mann-Kendall method for assessing trends in meteorological data [65]. Since then, several works have been drawn on meteorological parameters using the MK test [10,31,66,67]. For that purpose, we have applied the Mann-Kendall test in order to examine the existence of trends in the temperature distribution over Guinea.

The MK test is based on the following steps [68]:
The statistic S is written by the equation:

$$S = \sum_{i=1}^{N-1}\sum_{j=i+1}^{N} sgn(x_j - x_i) = 1, \tag{4}$$

where $N$ is the number of data points and $xj$ and $xi$ are data values at time $j$ and $i$ ($j>i$), respectively.

The statistics represent the number of positive differences minus the number of negative differences for all the differences considered.

$$\delta = (x_j - x_i), \tag{5}$$

$$sgn(\delta) = \begin{cases} 1 & \text{if } \delta > 0 \\ 0 & \text{if } \delta = 0, \\ -1 & \text{if } \delta < 0 \end{cases} \tag{6}$$

For large samples ($N>10$), the sampling distribution of $S$ is assumed to be normally distributed with the zero mean and variance as follows:

$$\text{var}(S) = \frac{N(N-1)(2N+5) - \sum_{k=1}^{n} t_k(t_k-1)(2t_k+5)}{18}, \tag{7}$$

where n is the number of tied (zero difference between compared values) groups, and $t_k$ is the number of data points in the $k$th tied group.

The Z-statistic or standard normal deviate is then computed by using the equation:

$$Z = \begin{cases} \frac{S-1}{\sqrt{var(S)}} & if S > 0 \\ 0 & if S = 0, \\ \frac{S+1}{\sqrt{var(S)}} & if S < 0 \end{cases} \tag{8}$$

where, if the computed value of $|Z| > 1.96$, then the null hypothesis of no trend is rejected.

In this study, the null hypothesis was tested at a 5% significance level. A positive value of $Z$ indicates an increasing trend, and a negative value of $Z$ indicates a decreasing trend.

For change-point detection in the trend, sequential Mann-Kendall (SQ-MK) test was proposed by Sneyers [68]. The main purpose of using this method is to determine the approximate time of beginning of a significant trend.

The SQ-MK test statistic is given by:

$$t_i = \sum_{j=1}^{j} n_i, \tag{9}$$

Where, for $x_i$ being the annual mean of the time-series, its magnitude (i=1,2,3,…..,n) is compared to $x_j$, (j = 1,2,3,…, j-1), and at each comparison, the number of case $x_i>x_j$ is counted and denoted by $n_i$.

Then, the mean of the statistic is calculated from:

$$E(t_i) = \frac{i(i-1)}{4} \tag{10}$$

The variance of the statistic is obtained by:

$$var(t_i) = \frac{i(i-1)(2i+5)}{72} \tag{11}$$

And finally, the sequential values are obtained from the below equation:

$$u(t_i) = \frac{t_i - E(t_i)}{\sqrt{var(t_i)}} \tag{12}$$

Similarly, the values of *u'*(*t*) are computed backwards, starting from the end of the series. The intersection of the curves showing forward (u) and backward (u') represents the time when the trend starts. The critical value for the 95% confident level is ± 1.96.

## 3. Results

### 3.1. Climatology of Temperature

Annual and Inter-Annual Variation of Temperature

A climatological analysis of the monthly average temperature variability at the 12 stations was performed. The annual variation of temperature is similar for the 12 stations, showing an annual and semi-annual cycle (see Figure 5a). The 12 stations are shared in three stations per natural region. The imposition of realistic surface wetness contrasts between the Sahara and equatorial Africa and leads to strong positive meridional temperature gradients at the surface and in the lower troposphere [69].

The inter-annual variation of the temperature reveals an increase in temperature (Figure 5b). The spatial distribution exhibits high temperatures in the Northern stations (Koundara, Siguiri and Kankan) and coastal stations (Conakry, Kindia and Boke) compared to the rest of the country. The specific case of the coastal region is linked to the combined effects of coast and topography. For the Northern stations, this temperature rise may be related to the positive south-north temperature gradient, except for stations in higher altitudes (e.g., Labe, which is the coldest site).

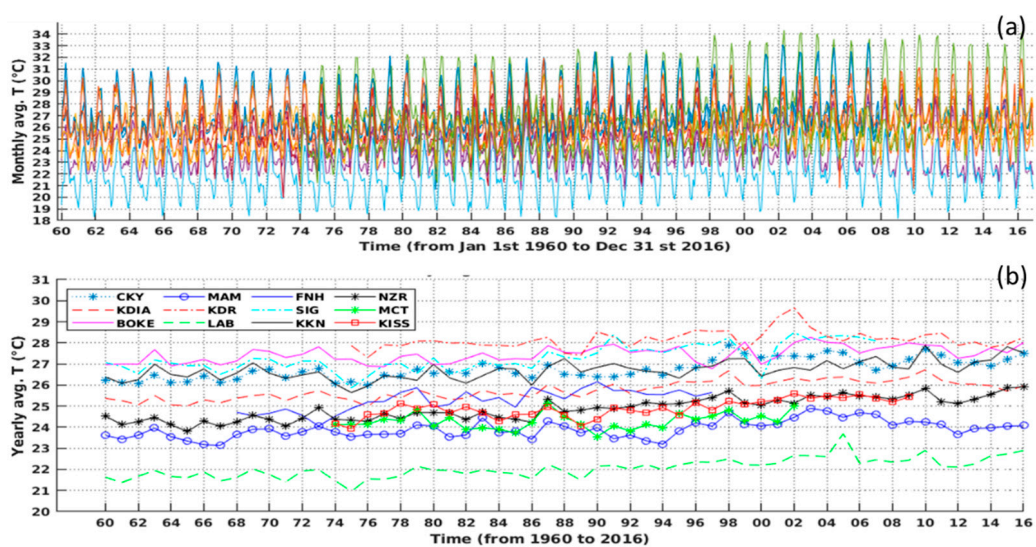

**Figure 5.** The inter-annual evolution of: (**a**) the monthly averaged temperature and (**b**) the yearly averaged temperature for the 12 weather stations in Guinea (CKY~Conakry, KDIA~Kindia, BOKE~Boke, MAM~Mamou, KDR~Koundara, LAB~Labe, FNH~Faranah, SIG~Siguiri, KKN~Kankan, NZR~N'zerekore, MCT~Macenta, and KISS~Kissidougou), from 1960 to 2016.

A wavelet transform analysis was used to identify the prevailing forcings that contribute to temperature variability. The Morlet wavelet allows a time-frequency decomposition giving a high-frequency resolution, because this wavelet is very localized from the point of view of frequencies. For more details on wavelet transformation, we suggest the reader to examine further references, such as Torrence and Compo [70], Bultheel [71] and Chang et al. [72], among others. The wavelet spectrum analysis showed energy concentrations in the six-months band and 12-months band at all the 12 sites. Figure 6a–d illustrates the semi-annual and annual cycles detected in the temperature variability at 4 sites (one by region) along the time period from 1960 to 2016.

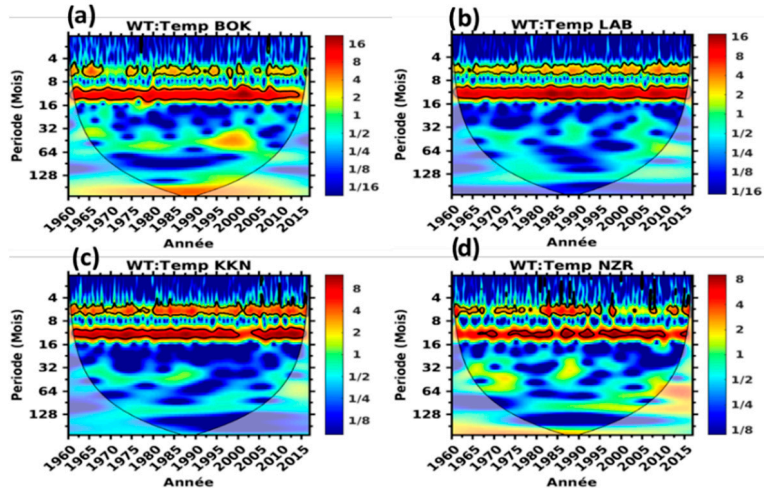

**Figure 6.** Dominant modes of variability detected by wavelet transform (WT) at 4 selected sites representative of each region: (**a**) Boke, (**b**) Labe, (**c**) Kankan and (**d**) N'zerekore.

According to the National Oceanic and Atmospheric Administration (NOAA), the term temperature anomaly means a departure from a reference value or long-term average. A positive anomaly indicates that the observed temperature was warmer than the reference value, while a negative anomaly indicates that the observed temperature was cooler than the reference value. Figure 5 shows the inter-annual evolution of temperature anomalies calculated according to the reference period of 1961–1990 and base period of more than 30 years [73]. The superimposed black line is the linear trend. As it is illustrated by Figure 7, we may notice that warming episodes appeared in the 1970s, 1984 and 1990s for most of the stations, and these warm periods were quite marked for the stations of the Northern and coastal regions. This implies that the 1970s, 1984 and 1990s droughts have been felt more in these regions compared to the South-Eastern area of the country. For an overall analysis, we note the persistence of positive anomalies after the 1990s. Although there are times when these anomalies are slightly reduced, the warming trend is still increasing across the country. Globally, this result is similar to the annual anomalies of the global average surface temperature calculated by the Climate Research Unit [74]. Salama et al. [75] found too an increase in the annual and monthly standardized temperature anomalies in the Tibetan Plateau.

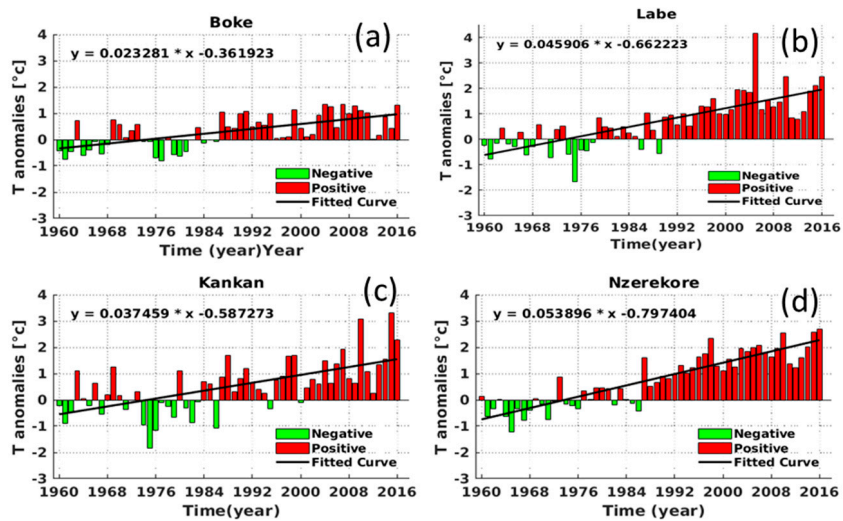

**Figure 7.** Inter-annual evolution of the surface temperature anomalies (°C) calculated for the station of (**a**) Boke in Lower-Guinea, (**b**) Labe in Middle-Guinea, (**c**) Kankan in Upper-Guinea and (**d**) N'zerekore in Forest Guinea from 1960 to 2016 (relative to the average temperature between 1961 and 1990). The superimposed curves (black line) denote the linear trend.

*3.2. Trends Estimation*

3.2.1. Temporal Evolution of Temperature and Simulation from the Trend-Run Model

Figure 8 shows the temporal evolution of the mean surface temperature values obtained from the stations of Lower-Guinea (Boke), Middle-Guinea (Labe), Upper-Guinea (Kankan) and Forest Guinea (N'zerekore) for the period from January 1960 to December 2016 described above. A similar upward trend in temperature is observed and characterized by both the annual and semi-annual oscillations pattern with the mean temperatures at the four stations (Figure 8a–d). The overall evolution of temperature shows upward trends (red lines). This result is in agreement with that found by Spear et al. [76], who found increases in the temperature of approximately 0.4 to 1.4°C along the coasts and other areas of Western Africa. The obtained upward trend in temperatures for the Lower-Guinea area stations is in agreement with earlier results found by Loua et al. [19].

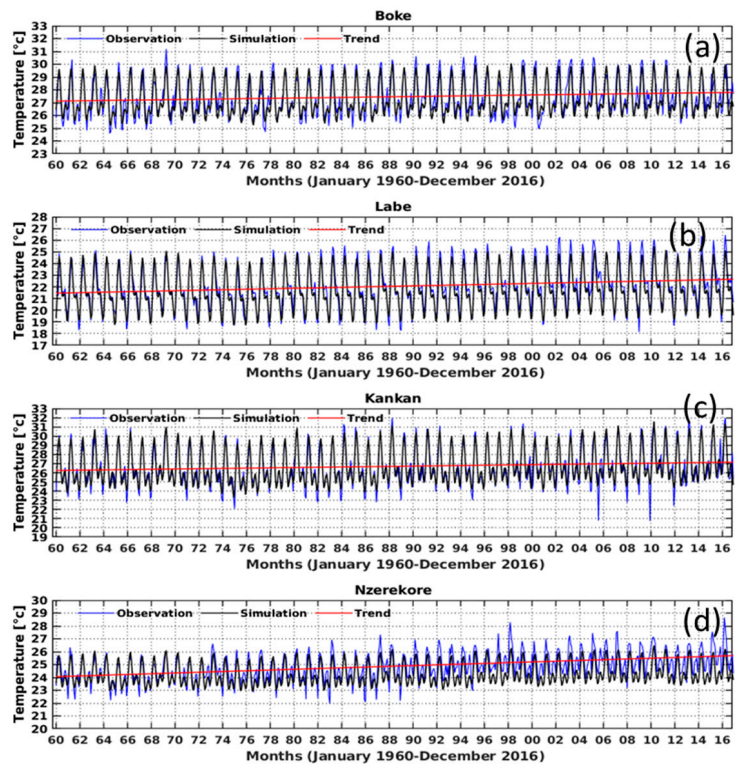

**Figure 8.** Temporal evolution of monthly surface temperature values (blue line) as observed over Guinea, superimposed by the simulated Trend-Run model (black line), while the straight red lines illustrate the obtained temperature trend at each weather station: (**a**) Boke, (**b**) Labe , (**c**) Kankan and (**d**) N'zerekore for the period from January 1960 to December 2016.

3.2.2. Contribution of Climate Forcings

The estimation of the different climate forcings' contributions included in the Trend-Run model with the corresponding standard deviation was done. The outputs from the model are shown in Table 2 with the corresponding values of the coefficient of determination ($R^2$). The model coefficient of determination values ranged between 0.60 (at N'zerekore) and 0.90 (at Siguiri), and the temperature trends per decade estimated by the Trend-Run model are reported in Table 2.

Results show that the AO component is the dominant forcing for most of the stations in Lower-Guinea (Conakry, Kindia and Boke); Middle-Guinea (Mamou, Labe and Koundara) and Macenta, Kissidougou and Faranah, but its contribution is the minimum at the Siguiri site (10.68% ± 0.11%) and the maximum at Boke (70.32% ± 0.27%). The AO minimum (maximum) contribution is found at Kankan ~11.48% ± 0.09% (at Boke ~70.32% ± 0.27%). The SAO contribution is higher at the Kankan site (47.60% ± 0.18%) and weak at the Boke site (15.05% ± 0.19%). Even if the SAO is a vital oscillation, particularly around the equator [77,78], this cannot explain the fact that the SAO contribution is dominant for the three stations mentioned above, because N'zerekore is the most Southern station. Thus, this may be attributed to other forcings that might be investigated.

In fact, the TNA appears as the third-most important forcing, and its maximum contribution value is obtained at the Siguiri site (5.16% ± 0.00%). Furthermore, the other forcings AN, Niño3.4 and SSN show weak contributions. The SSN contributes weakly (0.42% ± 0.03% as a maximum value at Faranah) to the temperature variation, while a natural link exists between the solar activity and climatic parameters, but it is not the cause of climate global warming [52]. The AN contribution is significant for the coastal and Southern stations compared to the rest of the country (0.88% ± 0.01% at Conakry as the maximum contribution).

The decadal trend values calculated by the Trend-Run model range between 0.04 °C ± 0.06 °C decade$^{-1}$ and 0.21 °C ± 0.06 °C decade$^{-1}$ throughout the country. This result is in agreement with the global analysis of the NOAA's merged land-ocean surface temperature anomalies reported by Vose et al. [79], who found a trend of 0.08 °C decade$^{-1}$ since 1901 and 0.16 °C decade$^{-1}$ since 1979. From the regional analysis, the decadal trend is higher in the coastal and Northern regions of the country and lowest over the forest and mountain regions (see Table 2). However, if each station is considered individually, the decadal trend maximum value is found at the N'zerekore site (0.21 °C ± 0.06 °C decade$^{-1}$), as compared to the rest of the stations (Table 2). As a significant finding, the model recorded high values of the coefficient of determination at both the stations in Upper-Guinea and Middle-Guinea. This suggests a relationship between the spatial distribution and model performance, revealing a good model performance in the coastal region, as shown by the Conakry $R^2$ = 0.69, Kindia $R^2$ = 0.82 and Boke $R^2$ = 0.89.

The observed trends in temperature are in agreement with those of some previous works, as Ringard et al [80], who explained warming in West Africa through the increase of temperature peaks and heatwaves. The northward warming found by our study is consistent with that found by Romes et al. [81], who showed a positive south-north gradient of trends in temperature and specified that meridional contrasts between the ocean and continent have increased significantly over the last 30 years.

**Table 2.** The corresponding values of geographical coordinates for the 12 weather stations. The coefficient of determination ($R^2$). Contribution and corresponding standard deviation (in percentages) of Annual Oscillation (AO), Semi Annual Oscillation (SAO), Tropical Northern Atlantic (TNA), El Niño 3.4 (Niño3.4), Sun Spot Number (SSN) and Atlantic Niño (AN) and decadal trend values, as obtained by the linear regression Trend-Run model at 12 weather stations of Guinea for the surface temperatures.

| Station | AO (%) | SAO (%) | TNA (%) | NIÑO 3.4 (%) | AN (%) | SSN (%) | $R^2$ | trend (°C/dec) |
|---|---|---|---|---|---|---|---|---|
| CKY | 32.15 ± 0.31 | 27.69 ± 0.29 | 2.85 ± 0.09 | 0.16 ± 0.02 | 0.88 ± 0.01 | 0.05 ± 0.05 | 0.69 | 0.14 |
| BOK | 70.32 ± 0.27 | 15.05 ± 0.19 | 0.39 ± 0.01 | 0.03 ± 0.01 | 0.63 ± 0.01 | 0.00 ± 0.04 | 0.89 | 0.14 |
| KIND | 51.19 ± 0.31 | 25.88 ± 0.14 | 0.14 ± 0.02 | 0.07 ± 0.01 | 1.02 ± 0.00 | 0.06 ± 0.03 | 0.82 | 0.07 |
| LAB | 56.10 ± 0.28 | 27.32 ± 0.20 | 1.23 ± 0.04 | 0.10 ± 0.01 | 0.16 ± 0.00 | 0.00 ± 0.01 | 0.88 | 0.12 |
| MAM | 50.81 ± 0.25 | 20.68 ± 0.16 | 0.40 ± 0.02 | 0.56 ± 0.03 | 0.22 ± 0.00 | 0.00 ± 0.02 | 0.81 | 0.06 |
| KDR | 66.38 ± 0.23 | 20.28 ± 0.12 | 0.14 ± 0.01 | 0.14 ± 0.01 | 0.14 ± 0.01 | 0.07 ± 0.01 | 0.87 | 0.13 |
| KKN | 11.48 ± 0.09 | 47.60 ± 0.18 | 1.43 ± 0.00 | 16.15 ± 0.11 | 0.01 ± 0.00 | 0.10 ± 0.00 | 0.87 | 0.09 |
| SIG | 10.68 ± 0.11 | 40.83 ± 0.22 | 5.16 ± 0.00 | 23.10 ± 0.16 | 0.21 ± 0.00 | 0.25 ± 0.00 | 0.90 | 0.17 |
| FNH | 48.22 ± 0.46 | 28.20 ± 0.34 | 0.02 ± 0.01 | 2.20 ± 0.10 | 0.23 ± 0.05 | 0.42 ± 0.03 | 0.79 | 0.11 |
| NZR | 23.20 ± 0.26 | 28.24 ± 0.29 | 1.35 ± 0.06 | 0.53 ± 0.04 | 0.69 ± 0.01 | 0.01 ± 0.05 | 0.60 | 0.21 |
| MCT | 38.44 ± 0.52 | 27.91 ± 0.44 | 2.41 ± 0.12 | 1.41 ± 0.10 | 0.84 ± 0.03 | 0.12 ± 0.07 | 0.72 | 0.04 |
| KISS | 34.22 ± 0.38 | 30.99 ± 0.37 | 0.95 ± 0.06 | 0.19 ± 0.03 | 0.24 ± 0.01 | 0.02 ± 0.03 | 0.69 | 0.13 |

In the first column of this table, CKY, BOK, KIND, LAB, MAM, KDR, KKN, SIG, FNH, NZR, MCT and KISS stand for Conakry, Boke, Kindia, Labe, Mamou, Koundara, Kankan, Siguiri, Faranah, N'zerekore, Macenta and Kissidougou, respectively.

### 3.2.3. Trend Estimation by MK and SQ-MK

Table 3 depicts a summary of the Mann–Kendall test model based on the yearly averaged temperature data. The power of the test as defined by the probability of rejecting *Ho* when the alternative hypothesis is the percentage of samples rejected by the test when a trend of a certain slope exists in the data [82].

The z-scores ($Z_1 = -1.96$, $Z_2 = 1.96$) and p-value ($\alpha = 0.05$) are well-known to indicate if there is a significant (or nonsignificant) downward (or an upward) trend in the time-series. Mainly, it was expected that significant positive z-scores calculated depicted a clearly upward trend over all the stations with positive z-scores ($> Z_2$) and p-values (far less than 0.05), except for the stations of Koundara and Mamou, where nonsignificant upward trends were found with p-values far greater than 0.05. To this end, the consequences of global warming are immediately perceptible in the Northern half of Guinea, as well as the coastal regions. This result is in agreement with the report of the national adaptation plan on climate change in Guinea (PANA, Plan d'Action National d'Adaptation aux Changements Climatiques) PANA [83].

However, the trend is exceptionally more significant in N'zerekore (z-score = 11.26). This may be related to the anthropogenic actions on the environment in Forest Guinea, especially in N'zerekore, because of anthropogenic activities like agriculture practices. Deforestation may also have influences on changing temperature trends at the local level. One could note that, in Forest Guinea, the rate of forest decline observed was 2.1% from 1981 to 2000, while, for the rest of the country, it was 0.5%, including mangroves [84].

**Table 3.** The corresponding values of geographical coordinates for the 12 weather stations. The available data number (n), the P-values and the Z-scores, as obtained by the linear regression Mann-Kendall trend test at 12 weather stations of Guinea for surface temperatures.

| STATION | LONG (°) | LAT (°) | N | P-VALUES | Z-SCORES | MK TEST CONCLUSION |
|---|---|---|---|---|---|---|
| | | | | Lower-Guinea | | |
| CONAKRY | −13.37 | 9.34 | 57 | $2.40 \times 10^{-01}$ | 9,45 | Sign. upward trend |
| BOKE | −14.18 | 10.56 | 57 | $4.90 \times 10^{-05}$ | 4.05 | -//- |
| KINDIA | −12.86 | 10.04 | 57 | $1.17 \times 10^{-07}$ | 5.29 | -//- |
| | | | | Middle-Guinea | | |
| LABE | −12.29 | 11.19 | 57 | $5.50 \times 10^{-08}$ | 5.43 | -//- |
| MAMOU | −10.08 | 10.38 | 57 | $1.15 \times 10^{-01}$ | 1.57 | Nonsign. upward trend |
| KOUNDARA | −13.31 | 12.34 | 42 | 0.02705 | 2.21 | Sign. upward trend |
| | | | | Upper-Guinea | | |
| KANKAN | −9.55 | 10.12 | 57 | $1.83 \times 10^{-4}$ | 3.74 | -//- |
| SIGUIRI | −9.37 | 11.74 | 41 | $4.03 \times 10^{-4}$ | 3.54 | -//- |
| FARANAH | −10.8 | 10.26 | 30 | $3.49 \times 10^{-05}$ | 4.14 | -//- |
| | | | | Forest Guinea | | |
| N'ZEREKORE | −8.83 | 7.75 | 57 | $2.20 \times 10^{-16}$ | 11.26 | -//- |
| MACENTA | −9.28 | 8.32 | 41 | 0.1706 | 1.4 | Nonsign. upward trend |
| KISSIDOUGOU | −10.11 | 9.19 | 36 | $3.41 \times 10^{-06}$ | 4.36 | Sign. upward trend |

Sign. stand by significant.

It is known that trend detection at the end of any time period applying the Mann–Kendall method does not give a complete trend picture (structure of the trend) for the whole series [85]. It is for this reason that the SQ-MK was performed for temperature change-point detection at all the 12 stations over the country. The results revealed a warming trend for all the stations except at the Macenta site (Figure 9(k)). In the coastal area, there is no turning-point change, because forward and backward trends do not intercept, which may mean that the recent increasing trend started a long time ago, and it may continue for a longer period in the future. Indeed, the trend was significant in 1973 for three stations (Figure 9(a–c)), and 1980, for the stations of Figure 9(a and c), and 1990 (Figure 9(b)) illustrate the progressive (red curve) and retrograde (blue curve) series.

In Middle-Guinea, there is no turning-point change for the station of Labe (Figure 9(d)), and the trend has stayed significant since 1990. However, the stations of Mamou (Figure 9(e)) revealed two trend changes in 1994 and 2003 and the station Koundara (Figure 9(f)) in 1988 and 1997.

In Upper-Guinea, the stations do not show any trend change in the corresponding study period, and the positive trend has been significant since around 1990 (Figure 9 (g–i)).

In Forest Guinea, there is no trend turning-point for the stations of N'zerekore and Kissidougou (Figure 9(j and l)) respectively. The positive trend has been significant at N'zerekore since around 1980; for the site of Kissidougou, the trend has been significant since the 1980s, and the actual positive trend significance started in 1998. The SQ-MK analysis for the station of Macenta (Figure 9(k)) indicates only one trend-turning point in 1995, but the trend is nonsignificant for the long-term annual temperature.

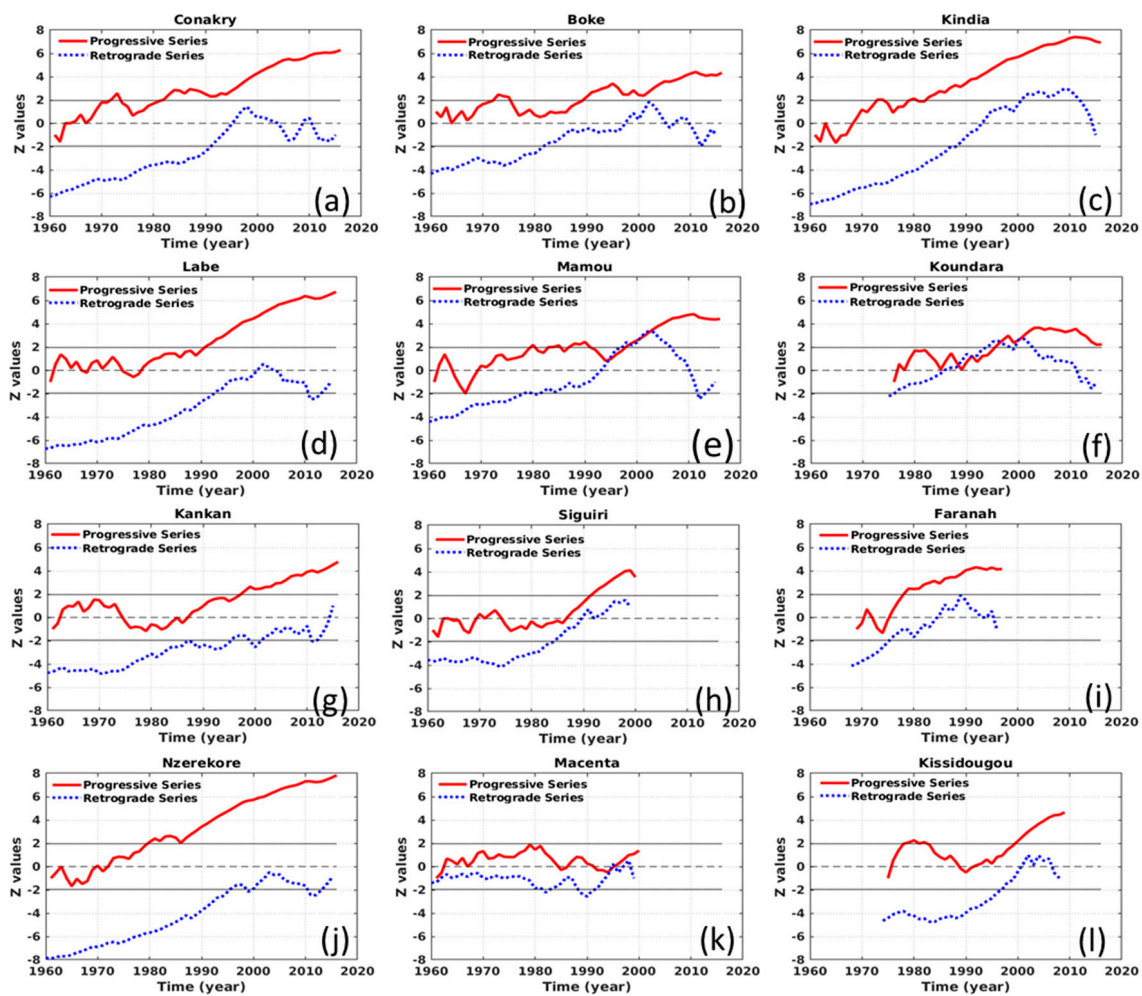

**Figure 9.** Sequential Mann-Kendall analysis of the annual trend of Lower-Guinea (**a–c**), Middle-Guinea (**d–f**), Upper-Guinea (**g–i**) and Forest Guinea (**j–l**). The red curves denote the forward sequential values of the statistic u(t), and the blues curves denote the backward sequential values of the statistics.

*3.3. Spatial Analysis*

Figure 10 shows the spatial seasonal variation of the surface temperature derived from MERRA 2 for the period 1980–2016. It is observed in this figure that the temperature is relatively higher for the coastal region, from September to February, presumably because of coastal effects. The maximum temperature is observed during March-April-May (MAM), and the Northern half of the country is warmer compared to the rest of the country. The first maximum appears in March and the second one in November at all the stations. By the start of the rain, Middle-Guinea and Forest Guinea are colder compared to the rest of the regions, presumably because of topography and vegetation, respectively.

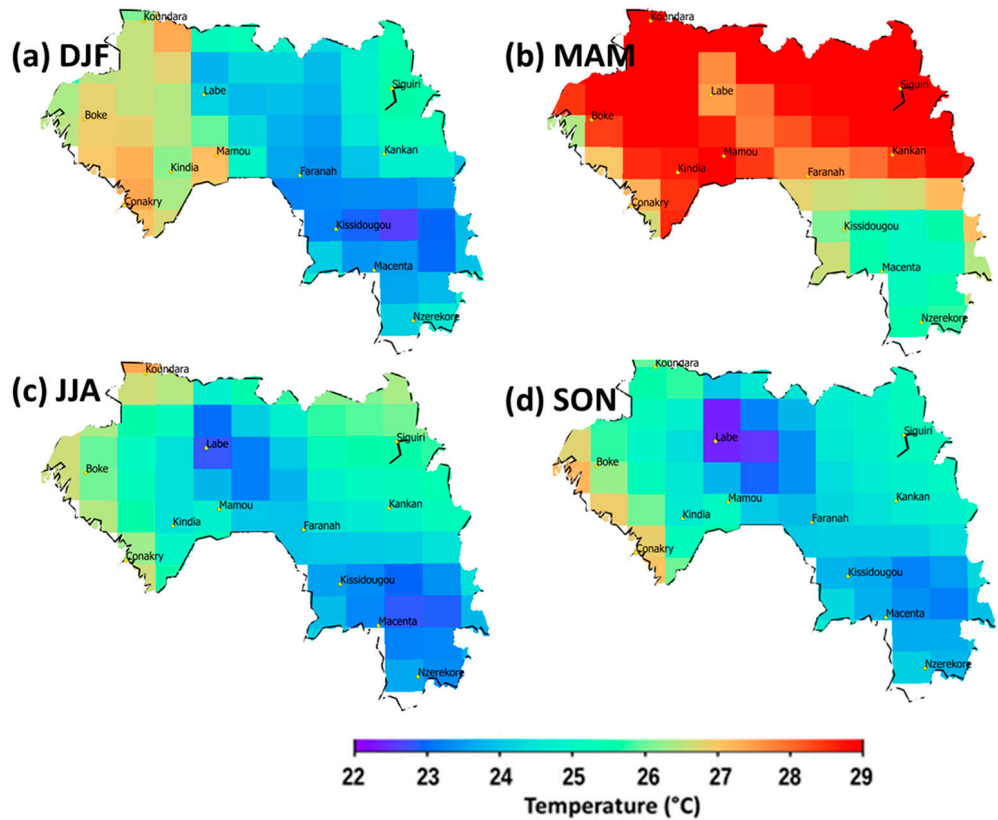

**Figure 10.** Seasonal variation of MERRA 2 surface air temperature over the 1980–2016 period: (**a**) December-February (DJF), (**b**) March-May (MAM), (**c**) June-August (JJA) and (**d**) September-November (SON).

The vegetation cover plays a significant role dictating the temperature of the area. Temperature is greater in areas which have low vegetation covers and lower in the dense forest areas. Thus, in this study, it is important to investigate the special dynamics of temperature over the study area. For this purpose, the Moderate Resolution Imaging Spectroradiometer (MODIS) dataset at a 250-m spatial resolution extending from 2000 to 2016 was used to calculate the normalized difference vegetation index (NDVI) for the study area (see Figure 11). In Figure 11, it can be observed that the land degradation is more prominent in the Northern half of the country. The Northern and the coastal regions of Guinea are concerned by mine exploitations. Large alterations to the surface land, such as cropland expansions, may change the surface temperature on local and regional scales. Pettorelli et al. [86] highlighted that, NDVI in ecological studies has recently rapidly increased and its possible key role in future research of environmental change in an ecosystem context allows the better predictions of the effect of global warming, biodiversity reduction or habitat degradation. In fact, He et al. [34] and Wang et al. [87] found a positive correlation between the temperature and NDVI early and late in the growing season and a weak correlation in the mid-growing season in Kansas during the 1989–1997 period. Southern areas surrounding stations such as Kissidougou, Macenta and N'zerekore have stronger NDVI values, and the surface temperatures of these areas are less sensitive to the seasonal warming during March-April-May (MAM) compared to the rest of the country.

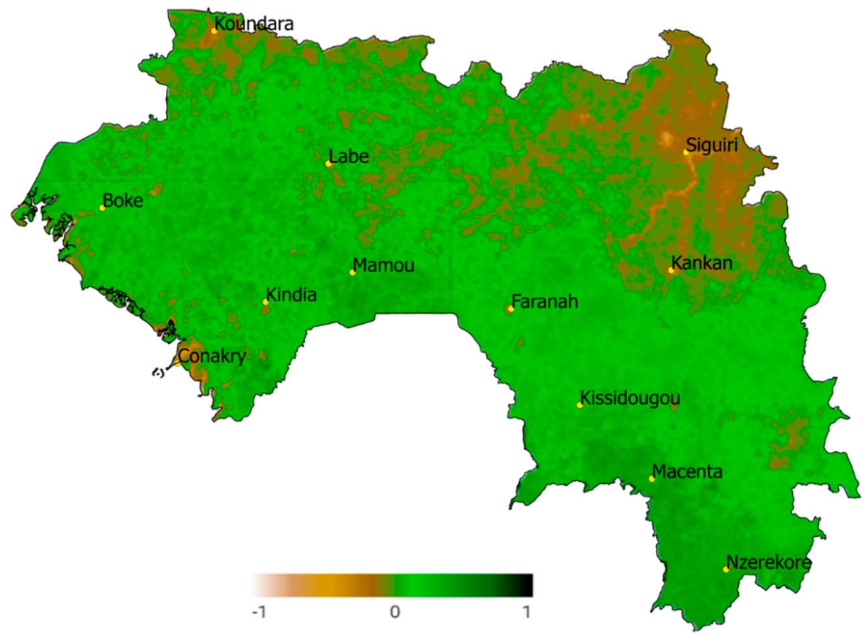

**Figure 11.** The averaged normalized difference vegetation index (NDVI) from 2002 to 2016 over Guinea.

## 4. Discussion

This analysis revealed an upward trend in temperature over all the sites concerned by the study from 1960 to 2016, except the Macenta station, where the upward trend was less significant. The peculiarity of this site may be linked to the presence of a protected forest in this region. The land use/land cover changes affect the local climate [32,33,88,89]. This result is in agreement with that found by Spear et al. [76], who found increases in the temperature of approximately 0.4 to 1.4°C along the coasts and other areas of Western Africa. Globally, this result is similar to the annual anomalies of global average surface temperatures calculated by the Climate Research Unit [74]. Salama et al. [75] also found an increase in annual and monthly standardized temperature anomalies in the Tibetan Plateau. The two almost different methods used in this study show a warming through the positive trends calculated. However, the spatial distribution of this warming is not uniform. Topography, land use and land cover are factors that influence the surface temperature (see Mahmood et al. [33]). The land alteration is more expanded in both the coastal and the Northern areas of Guinea because of the combined effects of mine exploitation and burning agriculture. According to the findings, the Northern half and the coastal zone of the country are the warmest. The observed trends in the temperature are in agreement with those of some previous works—for example, the northward warming found by our study is consistent with that found by Romes et al. [81], who showed a positive south-north gradient of trends in temperature and specified that meridional contrasts between the ocean and continent have increased significantly over the last 30 years. Ringard et al. [80] demonstrated the warming in West Africa through the increase of temperature peaks and heatwaves. Indeed, the obtained upward trend in temperature at the N'zerekore site was highlighted by Loua et al. [19].

According to this study, the topography, coastal-urban effects and land cover affects the temperature variation for some stations; thus, these local conditions might be examined for future works. Diba et al. [6] demonstrated in his work a strong contribution of the vegetation cover in a temperature decrease in Sahel. Comparatively to global warming, our results are in good agreement with the results of Wang et al. [1], who highlighted that, for the period of 1901–2010, the warming trend was found to be 0.109 °C decade$^{-1}$ with 95% confidence intervals between 0.081 °C and 0.137 °C. For the West Africa region, the observations already indicate an average increase in temperature of between 0.2 and 0.8 °C; when projected, this increases further to between 3.0 and 4.0 °C [84]. Moreover, Du et al. [90] highlighted that the variability of the diurnal temperature range is consistent with the variability of the solar flux from monthly to decadal time scales. Conversely, Jimmy et al. [10] found decreasing trends in the temperatures for some weather stations in South Africa.

However, this trend is exceptionally more significant in N'zerekore (z-score = 7.8); this means that the temperature at N'zerekore has the strongest trend but a lower coefficient of determination. It seems that this is mainly due to strong summer warming peaks in recent years that did not exist in previous years. This warming condition is linked too to anthropogenic actions within the environment in Forest Guinea, especially in N'zerekore, because anthropogenic activities like agricultural practices and deforestation may also have influences on changing temperature trends at the local level. In Forest Guinea, the rate of forest decline observed was 2.1% from 1981 to 2000, while, for the rest of the country, it was 0.5%, including mangrove forests [83]. Similar results were shown by Neumann et al. [91], who found clear positive trends with high levels of significance for a temperature time-series in the Volta Basin in West Africa. To compare this result to those of other neighbouring areas, and contrary to Diba et al. [6], Christidis et al. [92] have also suggested the emergence of warming since the 1980s in Central Sahel.

## 5. Conclusions

The aim of using the Trend-Run model, MK test and SQ-MK test was, firstly, based, on the one hand, to examine the existence of trends in the monthly temperature variation and, on the second hand, to quantify the contribution of various geophysical forcings. Secondly, to detect monotonous trends in annual temperature series and, besides indicating tendencies, to determine the approximate starting time of significant trends in the annual temperature series. All the methods used revealed an increase in the temperature trends, which could have direct consequences, such as increased evaporation during the drier periods, that can subsequently impact the available water resources, agriculture and health. The two methods are in good agreement on the spatial distribution of the temperature trends (see Appendix A). Additionally, the AO, followed by the SAO and TNA, were found to be the dominant contributors to the variability observed in the mean temperature datasets and the underlying AN, SSN and Niño3.4 may have weak influences on the variability of thermal structures and trend calculations. These results highlight that the Trend-Run model is highly applicable in all provinces across Guinea. Guinean's climate is changing, which means that long-term averages of the climate no longer provide a reliable guide to the future. Planning and decision-making needs to consider recent trends in the climate, as well as projected future changes, particularly where community assets or safety are concerned. It could help to make plans for stabilizing environmental conditions and building the capacity of weather forecasting services to respond to the agriculture, water resources, environment and climatic risks management in the future. Research needs to be especially encouraged for our study area (Guinea) because of its environmental strategic localization in West Africa and because of several ongoing mining projects in the country. Future work is planned to extend the Trend-Run model to include meteorological data (humidity and dewpoint), as well as additional explanatory parameters data, such as the ozone and normalized difference vegetation index.

**Author Contributions:** Conceptualization, R.T.L.; data curation, R.T.L.; formal analysis, R.T.L.; methodology, R.T.L., H.B., N.M. and N.B.; resources, H.B. and S.V.; software, R.T.L., H.B., N.B. and N.M.; supervision, H.B. and N.B. and validation, H.B., N.B., N.M., T.P., A.H., S.V. and Z.B. All authors have read and agreed to the published version of the manuscript.

**Funding:** This research received no external funding.

**Acknowledgments:** This work was undertaken in the framework of the French South-African International Research Group IRP-ARSAIO (International Research Project—Atmospheric Research in Southern Africa and the Indian Ocean) supported by the NRF and CNRS and by the Protea program (project No. 42470VA). The authors are thankful to the Guinean Weather Service for providing the essential data required for this study and to the NOAA web teams for providing climate indices. Authors are thankful to Paulene Govender for proofreading and to Souwala Dore for providing the photographs of the sites. We are also thankful to the anonymous reviewers for their insightful comments.

**Conflicts of Interest:** The authors declare no conflicts of interest.

**Appendix A**

Figure A1 shows the photographs of some observing sites of the Guinean meteorological observing network.

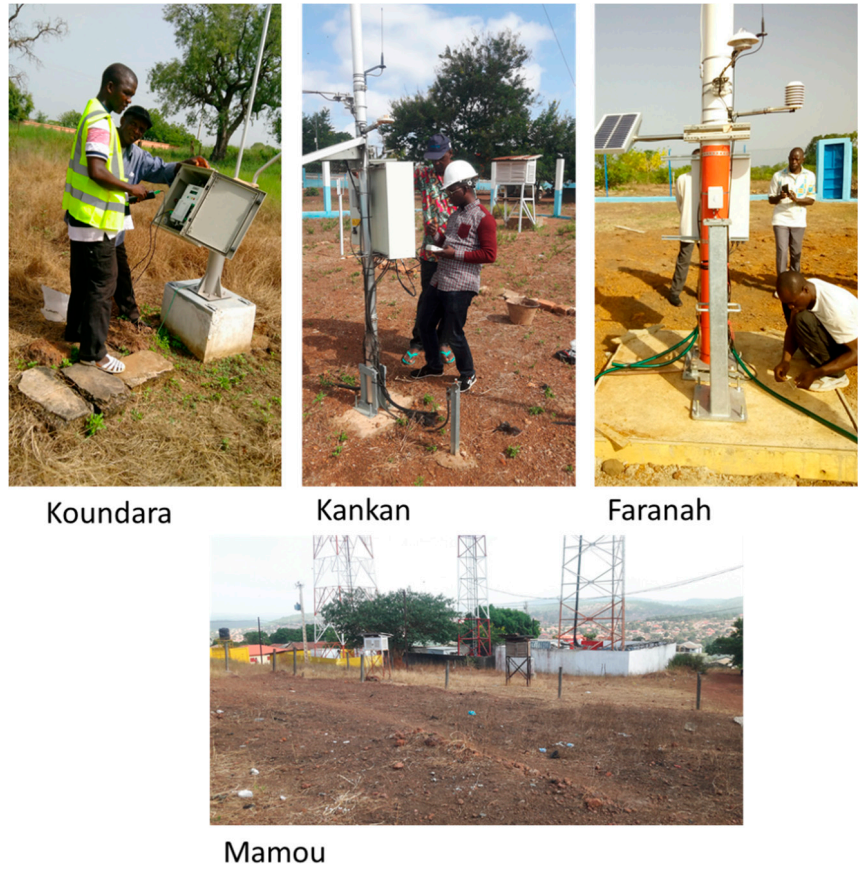

**Figure A1.** Photographs of observing sites (Koundara in the Northwest, Kankan in the Northeast, Faranah in the Center and Mamou in the Southwest Central regions of the country).

The spatial distribution of the temperature trend calculated by the two methods is depicted by the Figure A2. There is a strong correlation between the Trend-Run (red curve) and Mann-Kendall (black curve).

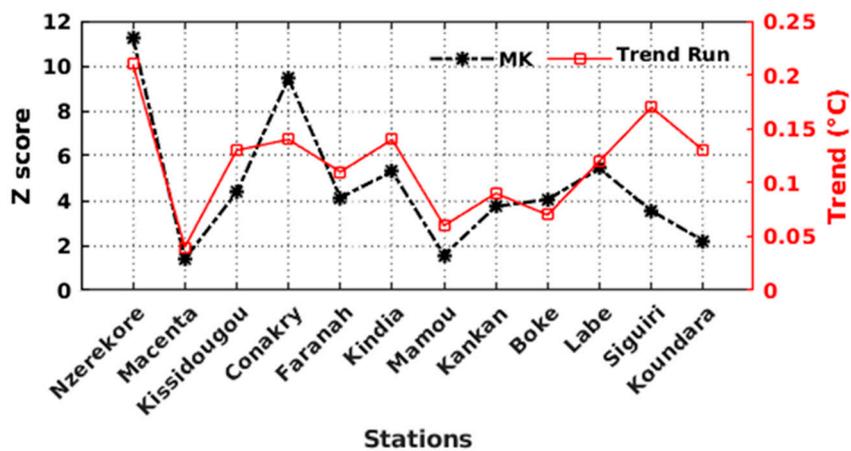

**Figure A2.** Spatial distribution of the temperature trend calculated by the Trend-Run (**a**) and MK (**b**).